\documentclass[12pt]{iopart}
\usepackage{amssymb}
\usepackage{epsf}

\begin{document}

\title[Two-dimensional Ising model in a random surface 
field]{Logarithmic corrections in the 
two-dimensional Ising model in a random surface field}

\author[M Pleimling \etal]{M Pleimling$^1$, F A Bagam\'ery$^{2,3}$,
 L Turban$^3$\ and F Igl\'oi$^{2,4}$}

\address{$^1$\ Institut f\"ur Theoretische Physik I, Universit\"at
Erlangen-N\"urnberg, D-91058 Erlangen, Germany}

\address{$^2$\ Institute of Theoretical Physics,
Szeged University, H-6720 Szeged, Hungary}

\address{$^3$\ Laboratoire de Physique des Mat\'eriaux, Universit\'e Henri 
Poincar\'e (Nancy 1), BP~239, F-54506 Vand\oe uvre l\`es Nancy Cedex, France}

\address{$^4$\ Research Institute for Solid State Physics and Optics,
H-1525 Budapest, P.O.Box 49, Hungary}

\ead{turban@lpm.u-nancy.fr}


\begin{abstract}
In the two-dimensional Ising model weak random surface field is predicted to be 
a 
marginally irrelevant perturbation at the critical point. We study this question 
by extensive Monte Carlo simulations for various strength of disorder. The 
calculated effective (temperature or size dependent) critical exponents fit with 
the field-theoretical results and can be interpreted in terms of the predicted 
logarithmic corrections to the pure system's critical behaviour.
\end{abstract}


\section{Introduction}

In an inhomogeneous system the local critical behaviour near 
localized or extended defects may differ considerably from the 
bulk critical behaviour in the regular lattice (for a review, 
see~\cite{igloi93}). One possible source of inhomogeneity is quenched 
(i.e., time-independent) randomness, which can be localized at 
the surface of the system (fluctuating surface coupling 
constants~[2--6],
microscopic terraces at the surface~\cite{selke02})
or at a grain boundary in the bulk of the system. It is known
experimentally~[8--10] that impurities may diffuse
from inside the sample and segregate on the surface or at 
grain boundaries. In adsorbed systems, quenched disorder is naturally 
present since adatoms may bind randomly on equivalent surface 
sites~\cite{cardy91}.

In a theoretical description of the local critical behaviour of 
these systems, close to the bulk critical point, one can use a 
coarse-grained picture in which quenched randomness couples to some 
local operator. The local operator considered in this paper 
is the surface order parameter, hence the perturbation is described by 
the introduction of random fields (RFs) localized at the surface. Usually
the RF has zero mean and its variance is used to characterize
the strength of disorder.

In the weak-disorder limit, the relevance or irrelevance of the
perturbation can be analyzed by making use of a Harris-type 
criterion~\cite{harris74}. The condition for the irrelevance
of RFs on a defect with dimension $d-1$ can be expressed 
in terms of the decay exponent for the local order parameter 
correlations in the pure system~\cite{diehl90a} as
\begin{equation} 
\eta_{\parallel} \ge 1\,.
\label{irrelevance}
\end{equation} 

A plane of RFs in the bulk often constitutes a relevant perturbation as 
it is the case for the two-dimensional (2D) Ising model with $\eta=1/4$.
Thus a new fixed point appears, which controls the local critical 
behaviour. This fixed point is expected to be a surface one since RFs 
tend to destroy the local order and the bulk defect then acts as an 
effective cut.

Surface RFs are irrelevant for the 3D Ising model as noted and
demonstrated through Monte Carlo (MC) simulations~\cite{mon88}. Curiously, in 
the 
case of a system with continuous symmetry, like the 3D Heisenberg model, surface 
RFs destroy the bulk long-range order~\cite{feldman02} according to Imry-Ma 
arguments~\cite{imry75}, although the perturbation is irrelevant at the ordinary 
surface transition according to~(\ref{irrelevance}). Among 2D systems the Ising 
model represents the borderline case, since $\eta_{\parallel}=1$~\cite{McCoyWu}.
For this model, field-theoretical investigations~\cite{cardy91,igloi91} predict 
that weak surface RF is a marginally irrelevant perturbation. Consequently, the 
surface critical properties of the random model are characterized by the 
critical 
singularities of the pure model supplemented by logarithmic corrections to 
scaling.

These theoretical predictions have not yet been confronted with the results of 
numerical calculations. In general, the observation and characterization of 
logarithmic corrections to scaling by numerical methods are notoriously 
difficult 
tasks, particularly in systems with quenched disorder. In this respect,
a well-known example is the diluted 2D ferromagnetic Ising model, for which the 
accurate form of the singularities was long debated~[18--23].

In this paper we present the results of a numerical study of the surface 
critical 
behaviour of the 2D Ising model in the presence of random surface fields. In 
section~2 we present the model and the known results about its critical 
properties. Then, through intensive Monte Carlo simulations, we determine 
effective surface magnetization exponents in two different ways. In section~3, 
they are obtained as a function of the deviation from the critical temperature. 
In section~4, we use a small homogeneous surface field at the critical point to 
deduce size-dependent exponents from the magnetization profiles. In section~5 we 
discuss the agreement between theoretical and numerical results.

\section{The model and its predicted surface critical properties}

We consider the Ising model on a $L\times M$ square lattice with the Hamiltonian
\begin{eqnarray}
&&{\cal H}=-J\sum_{i=1}^{L-1}\sum_{j=1}^M
(s_{i,j}s_{i+1,j}+s_{i,j}s_{i,j+1})
-\sum_{j=1}^M(h_1(j)s_{1,j}+h_L(j)s_{L,j})
\nonumber\\
&&s_{i,M+1}=s_{i,1}\,,\qquad
h_i(j)=\left\{
\begin{array}{ll}
h_{\rm s}+h & \mbox{with probability}\quad p=1/2\\
h_{\rm s}-h & \mbox{with probability}\quad p=1/2
\end{array}
\right.
\label{hamilt}
\end{eqnarray}
where $s_{i,j}=\pm 1$. $J$ is the first-neighbour exchange interaction. The RF 
$h_{\rm s}\pm h$ acts on the surface spins in the columns at $i=1$ and $i=L$ and 
periodic boundary conditions are used in the vertical direction. Our main 
interest is to calculate the averaged magnetization per column, $m_i=\langle 
|\sum_j s_{i,j}| \rangle / M$.

For the pure system, i.e. with vanishing surface field, the surface 
magnetization, $m_{\rm s}=m_1=m_L$ is exactly known in the thermodynamic limit 
($L,M \to \infty$)\cite{McCoyWu}:
\begin{equation}
m_{\rm s, pure} = \left[ \coth(2K)\, \frac{\sinh(2K) - 1}{\cosh(2K) - 
1}\right]^{1/2}\;,
\label{ms_pure}
\end{equation}
in terms of $K=J/k_{\rm B}T$, where $T$ is the temperature. At the critical 
point 
with $\sinh(2K_{\rm c})=1$ the surface magnetization vanishes as
\begin{equation}
m_{\rm s, pure} \approx m_0 t^{1/2}\;,
\label{ms_pure1}
\end{equation}
in terms of the reduced temperature, $t=(T_{\rm c}-T)/T_{\rm c}$, and 
$m_0^2=4(\sqrt{2}+1) K_{\rm c} \ln(2K_{\rm c})$.
The relevant length scale is the bulk correlation length which, for $t>0$, is 
given by~\cite{baxter}
\begin{equation}
\xi=\left[2 \ln(\sinh(2K))\right]^{-1}\;,
\label{xi}
\end{equation}
with the lattice constant for unit length. The correlation length diverges at 
the 
critical point as $\xi\approx [2\sqrt{2} K_{\rm c} t]^{-1}$. Thus the reduced 
temperature and the length scale are related by 
\begin{equation}
-\ln t \simeq .913 + \ln \xi\;.
\label{t_L}
\end{equation}
In the presence of random surface fields there are no exact results available. 
In 
this case one can use the replica trick to transform the semi-infinite system 
with a random surface field into $n$ semi-infinite replicas, coupled two-by-two 
through their surface spins, via nearest-neighbour interactions proportional to 
$h^2$. The average properties of the random system are obtained in the limit $n 
\to 0$. 

The surface critical properties of the system have been studied via two 
different 
methods, both using the differential renormalization group (RG) techniques
where the lengths are rescaled by a factor $\e^{l}$. The surface coupling 
between 
the replicas transforms as
\begin{equation}
h^2(l)=\frac{h^2}{1+\kappa h^2 l}\;.
\label{rg1}
\end{equation}

In the first approach~\cite{igloi91} a conformal mapping is used at the bulk 
critical point, with $h_{\rm s}=0$, to transform the $n$ semi-infinite replicas 
into $n$ infinite strips with width $L$, which are coupled to each other at both 
surfaces through $h^2$. The behaviour of the inverse correlation length is 
studied using degenerate perturbation theory to second order in $h^2$. From the 
transformation of the inverse correlation length on the strips under rescaling 
by 
$\e^{l}=L$, with $n=0$, and using the gap-exponent relation~\cite{cardy84}, one 
can identify the $L$-dependent, effective decay exponent $\eta_{\parallel}$ 
which 
is given by
\begin{equation}
\eta_{\parallel} =1+\frac{1}{\ln L}\;,
\label{eta}
\end{equation}
to leading logarithmic order.

In the second approach~\cite{cardy91} the behaviour under rescaling of the 
homogeneous part of the surface field is determined as
\begin{equation}
h_{\rm s}(l)=\frac{h_{\rm s} \e^{l/2}}{(1+\kappa h^2 l)^{1/2}}\;.
\label{rg2}
\end{equation}
The surface free energy density transforms as 
\begin{equation}
f_{\rm s}(t,h_{\rm s},h^2)=\e^{-l}f_{\rm s}[\e^{l/\nu}t,h_{\rm s}(l),h^2(l)]\;,
\label{fs}
\end{equation}
where $\nu$ is the correlation length exponent. Using~(\ref{rg2}), the surface 
magnetization reads 
\begin{equation}
m_{\rm s}(t,h^2)=\left.\frac{\partial f_{\rm s}}{\partial h_{\rm 
s}}\right|_{h_{\rm s}=0} 
=\frac{\e^{-l/2}}{(1+\kappa h^2 l)^{1/2}}\, m_{\rm s}[\e^{l/\nu}t,h^2(l)]\;.
\label{msl}
\end{equation}
With $\e^{l}=\xi\sim t^{-\nu}$ and $\nu=1$ according to~(\ref{xi}), 
ignoring higher order corrections, one obtains
\begin{equation}
m_{\rm s}(t,h^2) \sim \frac{t^{1/2}}{(1-\kappa_1 h^2 \ln t)^{1/2}}\;,\quad 0<t \ll 
1\;.
\label{ms}
\end{equation}
Thus the critical singularity of the pure model is supplemented by a logarithmic
corrections to scaling. From a practical point of view, one can define 
temperature-dependent effective exponents through
\begin{equation}
\beta_{\rm s}(t)=\frac{\ln[m_{\rm s}(t(1+\delta))/m_{\rm
s}(t(1-\delta))]}{\ln[(1+\delta)/(1-\delta)]}=\frac{1}{2}\left(1+\frac{1}{|\ln 
t|}+\cdots\right)\,,
\label{betas}
\end{equation}
with $\delta \to 0$. The last expression gives the leading logarithmic 
correction
following from equation~(\ref{ms}).

Taking into account the scaling relation $\eta_{\parallel} = 2 \beta_{\rm 
s}/\nu$ with $\nu=1$, the effective exponents in 
equations~(\ref{betas}) and~(\ref{eta}) correspond in terms of the relevant 
length scales, $L \rightarrow \xi \sim 1/t$.

The following two sections are devoted to a numerical test of the validity of 
these theoretical results.

\section{Effective exponents}

The surface critical exponents are deduced from the temperature dependence of 
the magnetization $m_i$ in the surface layers (for a review, 
see~\cite{pleimling04}). We set the 
homogeneous surface field to zero, $h_{\rm s}=0$, the strength of the random 
surface field ranging from $h=0.6$ to $h=1.5$, and take a finite reduced 
temperature, $t>0$. Systems of square and rectangular shapes
containing $L \times M$ spins, with $L$ and $M$ ranging from 50 to 1000, have been 
studied using the standard single-spin-flip method. Although systems with 
rectangular shapes ($M<L$) lead to reduced finite-size effects, the fraction of 
surface spins is smaller than for a square system and more runs are needed to 
achieve the same accuracy for the surface magnetization which is self-averaging. 
Thus we worked with square systems to spare computer time. 
The final data are obtained after averaging over at least 1000 different runs with 
different realizations of the random surface field. For every run, time average has 
been taken over a few $10^4$ Monte Carlo steps per spin after equilibration.

\begin{figure}[tbh]
\vglue3mm
\epsfxsize=9cm
\begin{center}
\mbox{\epsfbox{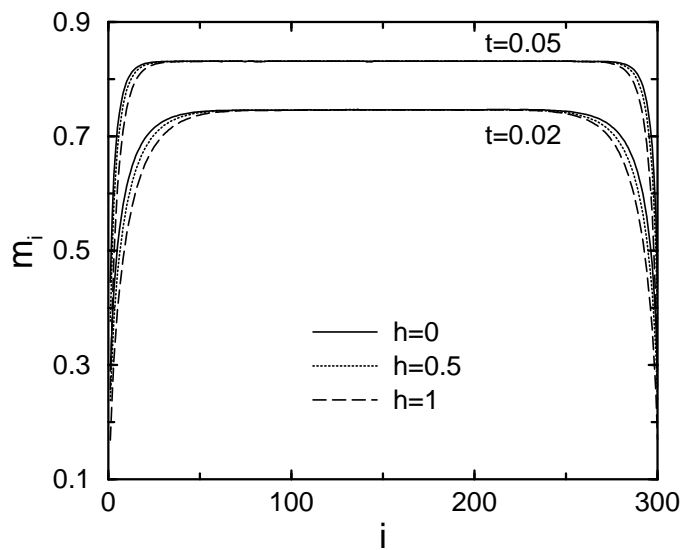}}
\end{center}
\vglue-3mm
\caption{Magnetization profiles with random surface fields $h=0$,
$0.5$ and $1$ at $t=0.05$ and $t=0.02$. The data have been obtained
for a system with $300 \times 300$ spins.}
\label{fig1}
\end{figure}

As an illustration we present $m_i$ at $t=0.05$ and $t=0.02$ in 
figure~\ref{fig1} for different strengths of the RF. The profile of the pure 
system is shown for comparison. For a given $t$ and different values of $h$, 
$m_i$ displays a plateau around $i=L/2$ for large enough systems. It corresponds 
to the bulk magnetization, $m_{\rm b}$, since its height is independent of $h$. 
If we approach the surfaces close enough, $i,L-i < \xi$, we enter in the surface 
region where the value of the magnetization is rapidly decreasing  to its 
surface 
value, $m_{\rm s}$. As seen in figure~\ref{fig1}, for a given $t$ the surface 
magnetization $m_{\rm s}$ and the inverse size of the surface region are 
decreasing with increasing disorder strength $h$. 

According to finite-size scaling, in a large but finite system, sufficiently
close to its critical point, $m_{\rm s}(t)$ behaves as 
$(-t)^{\beta_{\rm s}}f(\xi/L)$, where the scaling function, $f(x)$, tends to a 
constant for small values of its argument. In the actual calculations, we 
approach the transition point only to such a distance that the finite-size 
effects remain negligible and effective surface exponents are calculated 
using~(\ref{betas}). In practice, finite-size effects have been circumvented by 
adjusting the size of the sample in a standard approach~\cite{pleimling98}.
For a given value of $t$, data obtained for different system sizes are compared. 
Away from the critical point these data agree as long as the correlation
length is less than the extent of the smaller system. Closer to $T_c$ the
correlation length increases and at some stage it gets comparable to the
size of the smaller system. Finite-size effects then show up by a characteristic
fast drop of the effective exponent~\cite{pleimling98}. The smaller system is 
then discarded and the procedure is continued with two system sizes which still 
yield identical data at that temperature. This approach is somehow cumbersome 
but assures that the final data are essentially free of finite-size effects.

\begin{figure}[tbh]
\vglue3mm
\epsfxsize=9cm
\begin{center}
\mbox{\epsfbox{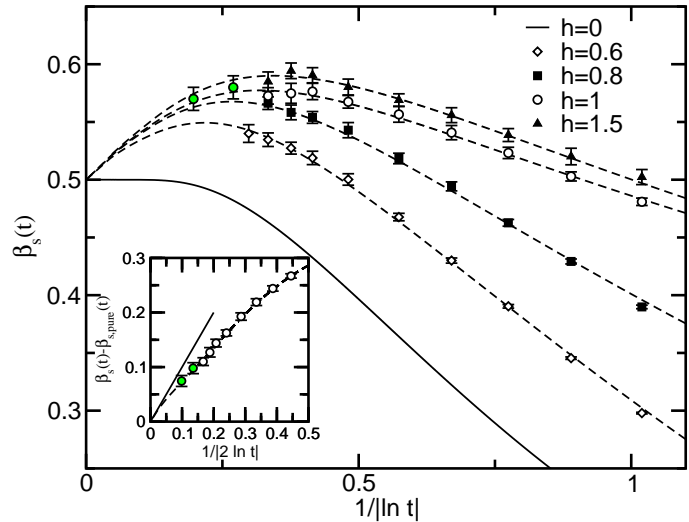}}
\end{center}
\vglue-3mm
\caption{Effective surface exponent for different strengths of the
random surface field. The two first points (grey circles) for $h=1$ were deduced 
from the short-distance 
behaviour of the critical profiles discussed in section~4. The broken lines 
fitting the numerical data
correspond to the formula in equation~(\protect\ref{pheno}),
with $a=-0.60$, $b=-0.23$ ($h=0.6$), $a=-0.53$, $b=-0.40$ ($h=0.8$),
$a=-0.40$, $b=-0.47$ ($h=1.0$) and $a=-0.47$, $b=-0.72$ ($h=1.5$). The
inset gives the difference between the effective exponents for the
random ($h=1$) and the pure systems. The straight line gives the
predicted leading logarithmic correction.}
\label{fig2}
\end{figure}

The effective exponents
are shown in figure~\ref{fig2} for different values of the strength 
of the RF. Here, in order to check the form of the logarithmic corrections in 
equations~(\ref{eta}) and~(\ref{ms}), $\beta_{\rm s}$ is plotted as a 
function of $1/|\ln t|$. For a given $t$ the effective exponents are 
increasing with $h$ as expected since RFs decrease local order. On the 
other hand, for a given $h$, $\beta_{\rm s}(t)$ first shows a monotonic 
increase when $t$ decreases and its value passes over the pure system's 
surface exponent, $\beta_{\rm s}=1/2$. Then, by further decreasing the 
reduced temperature, $\beta_{\rm s}(t)$ seems to approach a maximum value.
This saturation effect is more evident for large values of $h$. Unfortunately, 
the size limitation did not allow us to approach the transition point close
enough to follow the predicted decrease of the effective exponent.

Here, in order to compare the numerical results with the theoretical predictions 
and to extrapolate our data to $t \to 0$, we use the following expression, 
\begin{equation}
 m_{\rm s}(t)=m_0 t^{1/2}\frac{1+a\, t}{(1+b \ln t )^{1/2}}\;,
\label{pheno}
\end{equation}
which contains the leading analytic correction which follows from 
equation~(\ref{ms_pure}) and the leading logarithmic corrections to the 
fixed-point singularity given by~(\ref{ms}). Note that the two corrections have
different signs and their competing effect results in the non-monotonic 
temperature dependence of the effective exponent, $\beta_{\rm s}(t)$. For a 
given 
RF, we have fitted the surface magnetization data to the form given 
in~(\ref{pheno}) with the amplitudes $a$ and $b$ as free parameters. From this, 
the effective surface magnetization exponent was calculated and used to 
extrapolate the data points in figure~\ref{fig2}. The data extrapolate to the 
value of the pure system $\beta_{\rm s, pure}=1/2$ after an ``overshooting
effect''. A quite similar tendency was observed in~\cite{selke90} for 
the bulk magnetization exponent of the 2D random bond Ising model. In 
our calculations, however, the maximum value is almost reached for $h 
\ge 1$, which was not possible for the random bond model.

In order to check the leading logarithmic correction to the effective 
exponent, we have calculated the difference of $\beta_{\rm s}(t)$ and its
value in the pure system, $\beta_{\rm s, pure}(t)$, as calculated 
from~(\ref{ms_pure}) via equation~(\ref{betas}). This difference, which is 
plotted in the inset of figure~\ref{fig2}, no longer contains the leading
analytic correction; therefore we expect to be able to compare it with
the theoretical prediction in equation~(\ref{betas}). As seen in the inset for 
the random surface field, $h=1$, the corrections are compatible with theory, 
although much larger systems are needed in order to reach the asymptotic regime.

\begin{figure}[hbt]
\vglue3mm
\epsfxsize=9cm
\begin{center}
\mbox{\epsfbox{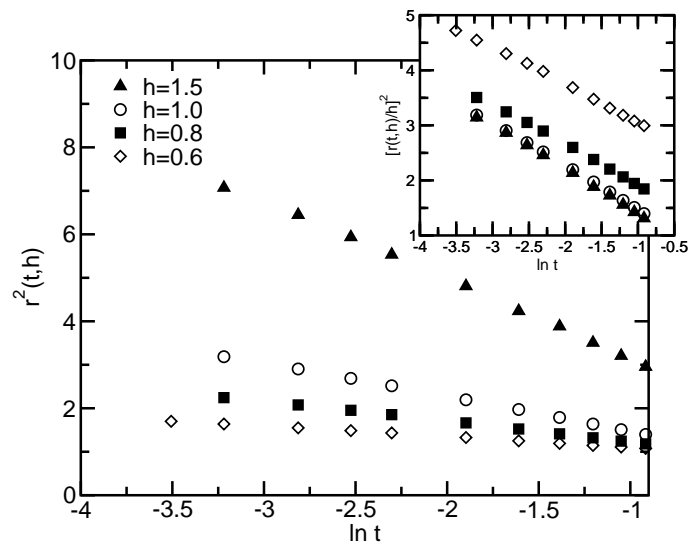}}
\end{center}
\vglue-3mm
\caption{Square of the ratio $r(t,h)$ of the surface magnetizations in the pure 
and in the disordered Ising model as a function of $\ln t $. The slopes are 
proportional to $h^2$ as shown in the inset. Error bars are much smaller than
the symbol sizes.}
\label{fig3}
\end{figure}

The form of the logarithmic corrections has been analyzed in still another way 
by 
forming the ratio, $r(t,h)=m_{\rm s, pure}/m_{\rm s}$, of the surface 
magnetizations in the pure and in the disordered systems. In this way the 
leading 
analytic correction to scaling is eliminated. As shown in figure~\ref{fig3}, the 
square of $r(t,h)$ has an asymptotic linear dependence on $\ln t$, the slope of
which is proportional to $h^2$, as shown in the inset of figure~\ref{fig3}. This 
is in complete agreement with the theoretical prediction in equation~(\ref{ms}).

\section{Critical profiles}

\begin{figure}[tbh]
\vglue3mm
\epsfxsize=9cm
\begin{center}
\mbox{\epsfbox{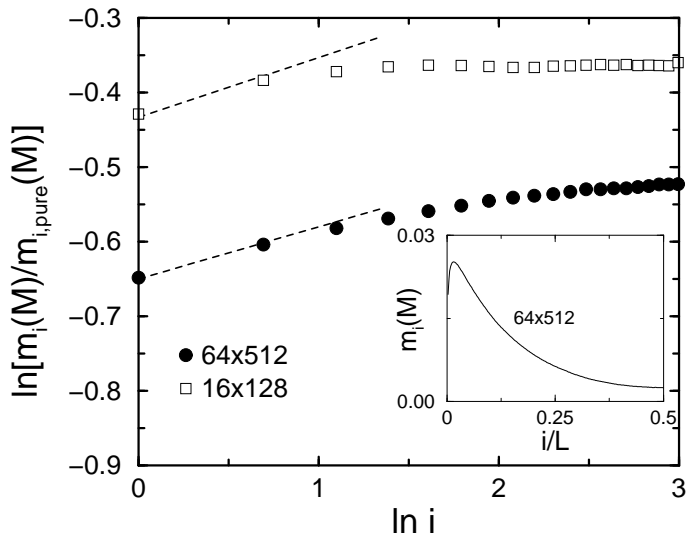}}
\end{center}
\vglue-3mm
\caption{Ratio of the initial part of the critical magnetization profiles 
calculated in the random and in the pure system in the presence of a small 
homogeneous magnetic field. The initial slope in the log-log plot corresponds to 
the difference in the effective magnetization exponents (see the text). Inset: one 
half of the magnetization profile in the random system.}
\label{fig4}
\end{figure}

In this section we study the system at the critical point, $t=0$; however, in the 
presence of a small homogeneous surface field, $h_{\rm s} \ll h$. A typical 
magnetization profile in the system with $M \ll L$ is shown in the inset of 
figure~\ref{fig4}. These data have been obtained with the Swendsen-Wang 
algorithm 
with a layer of ghost spins next to the surface~\cite{cr}. At least 1000 runs 
with different realizations of the random surface field were performed, the time 
average of every run resulting from typically $3 \cdot 10^5$ MC updates. 

As known 
from an analysis of the non-random system~\cite{cr} the profile first increases 
close to the surface and then decreases in the bulk. The surface critical 
exponent, $\beta_{\rm s}$, influences the form of the initial part and can be 
extracted from it. The non-vanishing surface field $h_{\rm s}$ introduces a new 
surface length scale, $l_{\rm s}$, which in 2D is given by 
$l_{\rm s}\sim h_{\rm s}^{-1/(1-\eta_{\parallel}/2)}$ and scales as 
$\sim h_{\rm s}^{-2}$ for the Ising model.

The initial part of the profile is restricted to $i \ll M, l_{\rm s}$ and, 
according to finite-size scaling theory it behaves as
\begin{equation}
m_i(M,l_{\rm s})=i^{-x_{\rm m}} g(i/M,i/l_{\rm s})\;,
\label{m_k}
\end{equation}
where $x_{\rm m}=\beta/\nu=1/8$ is the bulk magnetization scaling dimension.
For small values of its arguments the scaling function $g$ is expected to 
factorize as $g_0(i/M) g_{\rm s}(i/l_{\rm s})$. For the pure 2D Ising model the 
second term is logarithmic~\cite{bariev}, $g_{\rm s}(i/l_{\rm s}) \sim 
\ln(i/l_{\rm s})  \sim \ln(i h_{\rm s}^2)$, which makes the numerical analysis 
difficult. For the Ising model with RF on the surface, the first term is 
expected to behave as
\begin{equation}
g_0(y) \sim \frac{y^{1/2}}{(1+\kappa h^2 \ln y)^{1/2}}\;.
\label{g_y}
\end{equation}
This form incorporates logarithmic corrections which, with $i=1$ and $M\sim 
\xi$, 
are in agreement with the form of the surface magnetization given in 
equation~(\ref{ms}). Now, in analogy with~(\ref{betas}), an effective surface 
magnetization exponent can be defined as
\begin{equation}
\frac{\beta_{\rm s}(y)}{\nu}
=\frac{\ln[g_0(y(1-\delta))/g_0(y(1+\delta))}{\ln[(1+\delta)/(1-\delta)]}\,
\label{betas_y}
\end{equation}
which, finally when $y\ll 1$, will be a function $\beta_{\rm s}(M)$ of the 
characteristic length of the problem. 

In order to get rid of the logarithmic factor, $g_{\rm s}(i/l_{\rm s})$, we have 
calculated the ratio $m_i(M)/m_{i, {\rm pure}}(M)$ of the initial profiles in 
the 
random and pure systems. Assuming that $g_{\rm s}$ has the same logarithmic
singularity for both, we arrive to the conclusion that the ratio of profiles 
scales with the difference of the effective exponents, 
$\Delta \beta_{\rm s}(M)=\beta_{\rm s}(M)-\beta_{\rm s,pure}(M)$.

In the actual calculation we set $h_{\rm s}=.01$ and $h=1$, and performed MC 
simulations on systems with sizes $M=16$ and $64$ and different aspect ratios 
$\alpha=L/M$. For $\alpha=4$ and $8$, the initial part of the profiles 
turned out to be indistinguishable. The ratios calculated for the largest
$\alpha$ are presented in figure~\ref{fig4} and, from the extrapolated initial 
slopes in a log-log plot, the differences of the effective exponents take the 
values $\Delta \beta_{\rm s}(16)=0.08(1)$ and $\Delta \beta_{\rm 
s}(64)=0.07(1)$.
In order to compare these estimates with the effective exponents obtained in 
section~3 with a finite $t$, we use the correspondence of equation~(\ref{t_L}) 
with $M=\xi$. The data points obtained in this way are inserted in 
figure~\ref{fig2}. They seem to fit very well with the predicted theoretical 
curve and are located in the descending part of the curve. Therefore this 
calculation gives further support to the theoretical results about the form of 
the logarithmic corrections.

\section{Discussion}

Marginally irrelevant operators are responsible for logarithmic corrections
to scaling, the form of which can be often predicted by field theory and 
conformal invariance. According to second-order perturbation theory, the random 
surface field is expected to be such a marginally irrelevant operator at the 
surface fixed point of the 2D Ising model. This conjecture, which is made on the 
basis of the replica trick and in the weak disorder limit, is confronted here 
with the results of extensive MC simulations for varying strength of the 
disorder. The calculated effective surface magnetization exponent, which depends 
either on the distance $t$ from the critical point or on the finite size $M$ of 
the critical system, varies with these parameters. Since this variation is 
non-monotonic, a direct extrapolation to the fixed-point values cannot be made 
from the data available on finite systems. However, the variation of the 
effective exponents is in good agreement with the theoretical form,
which contains the predicted logarithmic correction to scaling to the pure 
systems critical behaviour. For the largest systems and for the strongest random 
fields, the numerical results are not too far from that obtained perturbatively 
in linear order. Therefore we interpret our results as numerical evidence in 
favour of the validity of the field-theoretical predictions.

\ack

We thank D E Feldman for bringing the results of~\cite{feldman02} to 
our attention.
FAB thanks the French Ministry of Foreign Affairs for a research grant. This 
work has been supported by the Hungarian National Research Fund under  grant no 
OTKA TO34183, TO37323, MO28418 and M36803, by the Ministry of Education under 
grant no FKFP 87/2001, by the EC Centre of Excellence (no~ICA1-CT-2000-70029).
Simulations have been done on the IA32 cluster of the Regionales Rechenzentrum 
Erlangen and at CINES Montpellier under project pnm2318. The Laboratoire de 
Physique des Mat\'eriaux is Unit\'e Mixte de Recherche CNRS~7556.


\Bibliography{99}

\bibitem{igloi93}
        Igl\'oi F, Peschel I and Turban L 1993 {\it Adv. Phys.} 
        {\bf 42} 683

\bibitem{diehl90a}
        Diehl H W and N\"usser A 1990 \ZP B {\bf 79} 69

\bibitem{diehl90b}
        Diehl H W and N\"usser A 1990 \ZP B {\bf 79} 79

\bibitem{pleimling98}        
        Pleimling M and Selke W 1998 {Eur. Phys. J.} B {\bf 1} 385 

\bibitem{diehl98}        
        Diehl H W 1998 {Eur. Phys. J.} B {\bf 1} 401

\bibitem{chung00}        
        Chung M-C, Kaulke M, Peschel I, Pleimling M and Selke W 2000 
        {Eur. Phys. J.} B {\bf 18} 655

\bibitem{selke02}
        Selke W, Pleimling M, Peschel I, Kaulke M, Chung M-C
         and Catrein D 2002 \JMMM {\bf 240} 349

\bibitem{weller85}
        Weller D, Alvarado S F, Gudat W, Schr\"{o}der K and 
        Campagna M 1985 \PRL {\bf 54} 1555

\bibitem{rau86}
        Rau C and Eichner S 1986 \PR B {\bf 34} 6347
        
\bibitem{rau87}        
        Rau C and Robert M 1987 \PRL {\bf 58} 2714  

\bibitem{cardy91}
        Cardy J 1991 \JPA {\bf 24} L1315

\bibitem{harris74}
        Harris A B 1974 \JPC {\bf 7} 1671

\bibitem{mon88}
        Mon K K and Nightingale M P 1988 \PR B {\bf 37} 3815

\bibitem{feldman02}
        Feldman D E and Vinokur V M 2002 \PRL {\bf 89} 227204

\bibitem{imry75}
        Imry Y and Ma S K 1975 \PRL {\bf 35} 1399

\bibitem{McCoyWu}
        McCoy B M and Wu T T 1973 {\it The Two-Dimensional Ising Model}
        (Cambridge: Harvard University Press) p 132

\bibitem{igloi91}
        Igl\'oi F, Turban L and Berche B 1991 \JPA {\bf 24} L1031 

\bibitem{dotsenko81}
        Dotsenko Vik S and Dotsenko Vl S 1981 {\it Sov. Phys.---JETP} {\bf 33} 37

\bibitem{shalaev84}
        Shalaev B N 1984 {Sov. Phys. Solid State} {\bf 26} 1811

\bibitem{selke90}
        Wang J-S, Selke W, Dotsenko Vl S and Andreichenko V B 1990
        {\it Physica} A {\bf 164} 221

\bibitem{kuhn94}
        K\"uhn R 1994 \PRL {\bf 73} 2268
        
\bibitem{plechko98} 
        Plechko V N 1998 \PL A {\bf 239} 289

\bibitem{shchur02}
        Shchur L N 2002 \PR E {\bf 65} 016107

\bibitem{baxter}
        Baxter R J 1982 {\it Exactly Solved Models in Statistical Mechanics} 
(London: Academic) p~118

\bibitem{cardy84}
        Cardy J 1984 \JPA {\bf 17} L385

\bibitem{pleimling04}
        Pleimling M 2004 \JPA {\bf 37} R79

\bibitem{cr}
        Czerner P, and Ritschel U 1997 Int. J. Mod. Phys. B {\bf 11} 2075

\bibitem{bariev}
        Bariev R Z 1988 Theo. Math. Phys. {\bf 77} 1090

\endbib
\end{document}